\documentclass[sigplan,10pt]{acmart}
\makeatletter
\global\@ACM@balancefalse
\makeatother
\renewcommand\footnotetextcopyrightpermission[1]{}
\usepackage{array}
\usepackage{algorithm}
\usepackage{algpseudocode}
\usepackage{mlir}
\usepackage{subcaption}
\usepackage{hyperref}
\usepackage{multirow}

\newcommand{\sysname}{Loom}
\newcommand{\sysop}[1]{\lstinline{loom.#1}}

\long\def\added#1{{#1}}
\long\def\modified#1{{#1}}

\title{Schedules Are Solvable Symbols: Tuning-Free Compilation of Tile Programs on Dataflow Architectures}
\author{Heru Wang}
\affiliation{%
  \institution{School of Computing, \\ National University of Singapore}
  \country{Singapore}
}
\email{heru.wang@nus.edu.sg}

\author{Wei Li}
\affiliation{%
  \institution{School of Computing, \\ National University of Singapore}
  \country{Singapore}
}
\email{liwei01@nus.edu.sg}

\author{Zhenyu Bai}
\authornote{Corresponding author.}
\affiliation{%
  \institution{School of Computing, \\ National University of Singapore}
  \country{Singapore}
}
\email{zhenyu.bai@nus.edu.sg}

\author{Tulika Mitra}
\affiliation{%
  \institution{School of Computing, \\ National University of Singapore}
  \country{Singapore}
}
\email{dcstm@nus.edu.sg}

\begin{abstract}
Modern AI and HPC accelerators increasingly expose \emph{dataflow features}: software-visible mechanisms for data movement and overlap, such as inter-core communication through the on-chip network and intra-core asynchronous pipelining. These features shift scheduling responsibility from hardware to the compiler, and---because placement, movement, and synchronization become software-visible---they also make the performance of static schedules predictable. Yet high performance on such hardware still relies on vendor-engineered kernel libraries or profile-based auto-tuning, whose embedded expert knowledge transfers poorly across architectures and algorithms.

We present {\sysname}, a tuning-free \emph{symbolic} compiler framework for tile-based SPMD programs on spatial dataflow architectures. The central idea is to treat tile-based SPMD compilation as a hardware-explicit static optimization problem. 
\modified{\sysname{} enumerates discrete spatial-mapping and communication candidates while keeping value parameters—such as tiling factors and pipeline knobs—symbolic within each candidate.}

From an explicit hardware description, it derives symbolic legality constraints and latency expressions, formulates one CP-SAT problem per schedule candidate, and jointly solves inter-core dataflow, intra-core asynchronous scheduling, and block sizes at compile time.

On two Tenstorrent generations, Wormhole and Blackhole, {\sysname} matches or exceeds the vendor-optimized TTNN library on GEMM, Flash Attention, and Flash Decode, out of the box and without per-shape profiling or profile-based platform-specific schedule tuning. These results suggest that hardware-derived symbolic compilation provides a retargetable alternative to profiling-based tuning for spatial dataflow architectures \added{while remaining interpretable by keeping optimization decisions traceable to source-level symbols.}
\end{abstract}
\begin{document}
\maketitle
\section{Introduction}
Modern AI and HPC kernels are increasingly dominated by structured tensor and stencil-style computations whose parallelism, control flow, and memory reuse are highly regular~\cite{ai-memory-wall}. For these kernels, moving data---not computing on it---has become the central performance and energy bottleneck~\cite{ai-memory-wall-2,memory-wall-tpu,horowitz}. Their regularity, however, also exposes static structure: tiles can in principle be placed in local memories, retained across uses, forwarded through the on-chip network, or overlapped with computation instead of being repeatedly fetched and scheduled dynamically. The problem this paper addresses is how a compiler can systematically exploit this static structure to plan data movement, rather than leaving it to dynamic hardware mechanisms or manual expert effort.

\modified{
Modern accelerators increasingly offload predictable control decisions to software, simplifying hardware control and reducing dynamic scheduling overhead while devoting more resources to bandwidth, local storage, and computation~\cite{emerging_accels}. We refer to the resulting software-visible mechanisms for orchestrating data movement and execution overlap as \emph{spatial dataflow features}. Across cores, these features support explicit on-chip communication, including forwarding and multicast, allowing data to be reused from local memories instead of repeatedly fetched from off-chip memory~\cite{plasticine,graphcore_poplar,ipu_solvers,cerebras_compiler_mach}. Within a core, they support asynchronous transfers, double buffering, and producer--consumer synchronization between memory and compute stages.

Such mechanisms are now common across spatial accelerators~\cite{tenstorrent_wormhole,cerebras_cs2,graphcore_ipu,sambanova_rda,sambanova_sn10,groq_tsm,groq_tsp,mtia,trainium,tesla_dojo}, while mainstream GPUs are moving in the same direction through asynchronous data movement, specialized pipelines, and increasingly explicit inter-SM communication. Collectively, these features allow software to orchestrate communication and computation around distributed local memories, improving data reuse and execution efficiency for regular workloads~\cite{ho_future_wires,benini_noc,maeri}.
}

\modified{Software has not kept pace with this trend. High-performance kernel development commonly follows one of two paths. The first relies on vendor libraries whose kernels are carefully optimized by hardware experts for a specific platform~\cite{cudnn,miopen}. These libraries perform well for supported operators, but their expert knowledge transfers poorly across architectures and algorithms, and users receive few tuning options.

The second uses tile-based SPMD DSLs, now a central abstraction for AI and HPC kernels~\cite{triton,tilelang,tilus,taichi,triton-shared}. Programmers write tile-level block programs, which the compiler and runtime lower to parallel threads. Because performance depends heavily on tile sizes and schedules, these systems often rely on profile-based auto-tuning or heuristic search~\cite{lewgpu,luthier,dsat}. This abstraction fits GPUs well: hardware dynamically manages the grid, L2 caches capture reuse across blocks, and warp switching hides intra-block latency. Consequently, existing compilers focus on efficient per-block code, leaving spatial-dataflow opportunities---core placement, NoC forwarding and multicast, and software-managed overlap---implicit and unexploited.
Thus, neither path provides portable, joint optimization of inter-core and intra-core dataflow: vendor libraries hard-code it for one platform, while tile-based DSLs do not expose it.}
Our key insight is that tile-based SPMD programs already contain the structure needed to plan both levels of dataflow, and that dataflow features make the resulting plans statically predictable. The grid is regular and homogeneous: it consists of many instances of the same block program over a predictable iteration space, which is exactly the information needed to plan inter-core dataflow---where block instances run, where their operands reside, and when data should be forwarded or reused through the on-chip fabric. The block program is heterogeneous: it describes the concrete sequence of loads, computation, stores, buffering, and synchronization needed by one tile of work, which is exactly the information needed to plan intra-core dataflow---asynchronous movement, double buffering, and producer--consumer overlap within each core. Because dataflow features expose placement, movement, and synchronization decisions to software, schedules planned this way become predictable to the compiler; with sufficient hardware information, good schedules can therefore be \emph{selected} statically rather than \emph{discovered} through profile-based tuning.

The challenge is that the two levels cannot be optimized independently. Inter-core choices determine which data arrive at each core, at what rate, and with what reuse opportunities; those choices constrain the local buffers, synchronization points, and pipeline stages available to the block program. Conversely, intra-core schedules determine how much local memory is required, how long data must be retained, and how much communication or computation can be overlapped, which changes the best grid placement and forwarding pattern. Exploiting SPMD for spatial dataflow therefore requires hardware information: local-memory capacity, network topology and bandwidth, off-chip bandwidth, synchronization costs, and compute throughput determine which combinations of grid-level and block-level decisions are legal and profitable.

Worse, this joint space contains many value-based design choices. SPMD block sizes, buffer allocations, forwarding groups, and pipeline depths are not small categorical options: they are integer parameters whose values change the amount of computation per block, local-memory footprint, communication volume, available parallelism, and overlap opportunities. Enumerating all combinations is impractical, but fixing these values too early obscures the interactions that determine whether a dataflow plan is legal and profitable. A useful compiler must therefore reason about these choices symbolically, using hardware constraints and analytical cost models to prune invalid mappings and compare candidate schedules without exhaustively compiling and profiling every instance.

In this work, we present \sysname{}, a tuning-free end-to-end compiler framework for tile-based SPMD programs on spatial dataflow architectures. The central idea is to treat tile-based SPMD compilation as a hardware-explicit static optimization problem: instead of optimizing a fixed tile configuration, \sysname{} keeps SPMD block sizes and pipeline decisions symbolic throughout the compilation flow, derives legality constraints and latency expressions from explicit hardware descriptions, and jointly optimizes inter-core dataflow, intra-core asynchronous scheduling, and SPMD block size decisions at compile time. By making hardware resources such as local memory capacity, memory bandwidth, and compute throughput visible to the optimizer, \sysname{} can reason about the correlations among these choices rather than relying on isolated, architecture-specific rules. Unlike auto-tuned DSL stacks, no per-kernel hardware profiling is performed in the loop; unlike vendor libraries, all hardware knowledge is parameterized in the architecture description, so retargeting a new hardware generation only requires updating that description rather than re-engineering kernels. Beyond selecting high-performance mappings, the solver exposes dominant bottlenecks in interpretable terms, providing actionable insight for both compiler engineers and hardware designers.

We evaluate \sysname{} on two generations of Tenstorrent hardware, Wormhole and Blackhole, using three representative operators: GEMM, Flash Attention, and Flash Decode.
Compared with the vendor-optimized TTNN library, \sysname{} matches or exceeds TTNN on GEMM (with geometric-mean speedups of $0.99\times$ and $1.16\times$ on Wormhole and Blackhole, respectively) and Flash Decode ($1.01\times$ and $1.17\times$), and achieves a $2.15\times$--$2.29\times$ geometric-mean speedup on Flash Attention. All results are obtained out of the box, showing that \sysname{}'s architecture-parameterized compilation flow retargets across hardware generations. The ablation studies further show that the solver usually selects spatial mappings and block sizes that are locally optimal or close to the best measured alternatives. Together, these results suggest that explicitly modeling dataflow choices, memory-hierarchy constraints, NoC communication, and intra-core pipelining enables a compiler to recover much of the expert scheduling knowledge traditionally embedded in hand-tuned vendor kernels. 
\added{The source code of \sysname{} is publicly available at \url{https://github.com/ecolab-nus/loom}.}

\begin{figure*}[htbp]
    \centering
    \includegraphics[width=\textwidth,keepaspectratio]{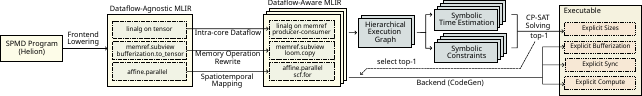}
    \caption{Overview of the \sysname{} framework.}
    \Description{Compilation flow from a symbolic SPMD kernel through MLIR transformations, schedule generation, analytical modeling, constraint solving, and target code generation.}
    \label{fig:framework}
\end{figure*}
\section{Framework}
\subsection{Overview}

\modified{Figure~\ref{fig:framework} summarizes \sysname{}'s end-to-end compilation flow. \sysname{} first lowers a tile-based SPMD program into a symbolic, \emph{dataflow-agnostic} MLIR (Multi-Level Intermediate Representation)~\cite{mlir} program. We call this representation dataflow-agnostic because it preserves the logical SPMD grid, block-level computation, and global-memory accesses, while leaving producer-consumer dataflow, local-memory usage, and buffer readiness semantics implicit. \sysname{} then reconstructs the producer-consumer structure and makes the local-memory and readiness semantics explicit, producing dataflow-aware MLIR while keeping their concrete realization abstract.
The architectural information in Figure 2 guides the remaining stages. 
Starting from the dataflow-aware MLIR, \sysname{} uses the target topology and NoC capabilities to enumerate alternative spatiotemporal mappings of the logical SPMD grid and corresponding data-movement strategies. Each combination forms a schedule candidate, while value parameters such as block sizes and pipeline choices remain symbolic. For each candidate, \sysname{} constructs a Hierarchical Execution Graph (HEG) that captures its program structure and potential overlap. \sysname{} then combines the HEG with the target resource and performance information to derive a symbolic latency expression and legality constraints, which together form a CP-SAT optimization problem. \sysname{} solves these candidate-specific problems, selects the feasible candidate with the lowest modeled latency, substitutes the solved symbolic values into the selected program, and lowers it through the target backend to the final executable.}

\begin{figure}[ht]
  \centering
  \includegraphics[width=\linewidth]{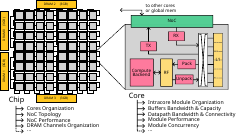}
  \caption{Architectural information used by \sysname{}.}
  \Description{Diagram of the hardware topology, communication capabilities, resource constraints, and performance models supplied to \sysname{}.}
  \label{fig:arch}
\end{figure}

\paragraph{Stage 1: Lowering SPMD programs to symbolic MLIR}
\modified{
Given a Helion SPMD program, \sysname{} lowers its logical grid, block-level control flow, tensor computation, and global-memory accesses into MLIR while preserving symbolic block-size parameters. The output is a dataflow-agnostic MLIR program that retains both grid-level parallelism and block-level computation but leaves producer-consumer data movement and synchronization implicit.
}

\paragraph{Stage 2: Reifying intra-core dataflow.}
\modified{
Given the dataflow-agnostic MLIR from Stage~1, \sysname{} converts tensor dependences represented by Static Single Assignment (SSA) use-def chains into explicit producer-consumer dependences over virtual buffers.
The output is a dataflow-aware MLIR program in which producer-consumer handoffs are explicit while their physical storage and synchronization resources remain abstract.
}

\paragraph{Stage 3: Concretizing virtual buffers.}
\modified{
\sysname{} uses liveness analysis to bind virtual buffers to physical local-memory and synchronization resources, which can be reused when legal.
The output replaces abstract virtual-buffer resources with memory references and synchronization operations.
}

\paragraph{Stage 4: Topology- and NoC-aware candidate generation.}
\modified{Given the symbolic logical SPMD grid and the target topology, \sysname{}
enumerates spatiotemporal mappings of program parallel axes onto the available
cores, with any remaining grid extent executed temporally.
For each mapping, \sysname{} analyzes the resulting global-memory accesses for
cross-core reuse and considers the inter-core transfer patterns supported by
the target NoC.
The combination of a spatiotemporal mapping and its data-movement choices forms
a schedule candidate with explicit spatial and temporal placement and inter-core
communication.}

\paragraph{Stage 5: Symbolic performance modeling and constraints.}
\modified{
For each schedule candidate, \sysname{} constructs an Hierarchical Execution Graph (HEG) that captures its structured execution and potential overlap. 
Combining the HEG with target performance and resource models yields a symbolic latency expression and legality constraints, forming one CP-SAT problem per candidate.
}

\paragraph{Stage 6: Solving and target code generation.}
\modified{
\sysname{} solves each candidate-specific CP-SAT problem independently, discards infeasible candidates, and selects the feasible candidate with the lowest modeled latency.
It then materializes the solved symbolic values in the selected MLIR program and lowers it through the hardware-specific backend to the final executable.
}

\subsection{Symbolic MLIR}
The standard MLIR infrastructure does not provide the kind of first-class compile-time symbol needed by \sysname{}: a value that has a stable source-level name, a finite domain, and can flow through the IR as an SSA value until the solver assigns it a constant. MLIR already uses the term ``symbol'' for named IR entities such as \lstinline{@foo}; in this paper, we call those entities \emph{labels} and use \emph{symbol} to mean a compile-time unknown introduced by our extension. Our symbolic MLIR extension keeps such unknowns explicit in shapes, loop bounds, index expressions, memory footprints, and performance-model inputs throughout the compilation pipeline.

A symbol is introduced by an \sysop{sym} operation:
\begin{lstlisting}[style=mlir-druvbox-light-high-inline,numbers=none]
%0 = (*@\texttt{loom}@*).sym @tile_b { upper_bound = 16 : index }
\end{lstlisting}

The operation has two roles. The SSA result, here \lstinline{

The label, here \lstinline{@tile_b}, is the human- and solver-facing identity of the same unknown. A label can originate from several sources, but it most commonly comes from a block-size symbol that already exists in the Helion kernel. For example, when a source loop such as \lstinline{for tile_m, tile_n in hl.tile([m, n]):} is lowered, \sysname{} creates symbols whose labels preserve the source names \lstinline{@tile_m} and \lstinline{@tile_n}. These labels are carried through the compilation pipeline and become variable names in the CP-SAT formulation.

This separation makes the optimized schedule interpretable. Compiler passes manipulate the SSA values when building symbolic bounds, access windows, buffer footprints, multicast regions, and HEG workloads. The modeling and solving layer uses the labels to name the corresponding CP-SAT variables and to report the solution back in source-level terms, such as the selected values of \lstinline{tile_m} and \lstinline{tile_n}. \sysname{} uses the same mechanism for symbols introduced later by the compiler, such as spatial tiling factors, and pipelining knobs, so all solved quantities remain traceable.

After CP-SAT solving, each \sysop{sym} is replaced by the solved constant, and all dependent expressions are rewritten accordingly. The output is a concrete MLIR kernel variant whose loop bounds, buffer sizes, communication regions, and backend choices are fixed, while the selected constants remain traceable to the original source-level symbol labels.

\subsection{Reifying Tensor Values into Synchronized Buffers \texorpdfstring{\sysop{vb}}{loom.vb}} \label{sec:vb}

\modified{At this point, the IR produced by Stage~1 represents block computation as MLIR \lstinline{linalg} operations over \lstinline{tensor} values, with producer-consumer dependences preserved through SSA use-def chains. 
These dependences are sufficient to preserve program semantics, but not to exploit the intra-core asynchronous execution exposed by modern dataflow architectures: memory and compute workers can proceed independently and form pipelines as long as their input data are ready. 
Exploiting this capability therefore requires the compiler to explicitly represent when local storage is writable and when produced data become readable. 
\sysname{} introduces \emph{virtual buffers} as a target-independent protocol for expressing this local storage and data-readiness state. 
Compiler passes use virtual buffers to coordinate producer-consumer handoff through readable and writable state, while later lowering stages choose the physical buffers, synchronization resources, and target instructions that implement the protocol.}

\sysname{} exposes the virtual buffer abstraction through a synchronization interface and a data-access interface.
The synchronization interface consists of four \sysop{vb.*} operations.
Before a consumer reads a value, \sysop{vb.acquire} waits until the corresponding virtual buffer contains the required amount of readable data; after the read completes, \sysop{vb.release} records that the consumer no longer needs that data.
Before a producer writes a value, \sysop{vb.reserve} waits until the destination virtual buffer has enough writable capacity; after the write completes, \sysop{vb.commit} makes the newly produced data readable by downstream consumers.
Together, these operations make the synchronization that governs access ordering explicit in the IR, while still deferring the choice of concrete storage and synchronization resources.

Conceptually, a virtual buffer behaves like a circular buffer: readers and writers use different views to access the buffer because their accesses are derived from the buffer's read and write positions, respectively.
This lets ordinary memory operations access the payload while the \sysop{vb.*} protocol maintains the corresponding readiness state.

\begin{figure}
    \centering
    \includegraphics[width=\columnwidth,keepaspectratio]{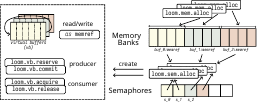}
    \caption{\sysname{}'s virtual-buffer abstraction and related primitives.}
    \Description{Diagram showing a virtual buffer bound to physical memory and semaphore resources and accessed through acquire, release, reserve, and commit operations.}
    \label{fig:memory-binding-example}
\end{figure}

Virtual buffers can be implemented differently across hardware targets, but each implementation must provide two capabilities: storage for the payload and a counting synchronization mechanism that tracks readable data and writable capacity.
As shown in Figure~\ref{fig:memory-binding-example}, \sysop{mem.alloc} allocates the required storage resource and \sysop{sem.alloc} allocates the synchronization resource; binding these resources materializes the virtual buffer for target lowering.
The target backend then lowers the resulting storage accesses and synchronization operations into target-specific instructions.

The tensor semantics in the dataflow-agnostic MLIR are reified into virtual-buffer semantics. Figure~\ref{fig:classic-df} illustrates the protocol for a matmul. Global-to-local copies reserve and commit \texttt{vb\_A} and \texttt{vb\_B}; the matmul acquires and releases these inputs while reserving and committing \texttt{vb\_C}; the final copy consumes \texttt{vb\_C} through the same protocol. Thus, data movement and computation use a uniform virtual-buffer interface.

\begin{figure}
  \centering
  \includegraphics[width=\linewidth]{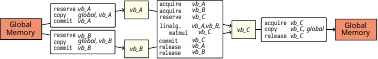}
  \caption{Typical intra-core dataflow.}
  \Description{Producer-consumer dataflow for global-to-local copies, matrix multiplication, and virtual-buffer synchronization within a core.}
  \label{fig:classic-df}
\end{figure}

\begin{figure}[ht]
  \centering
  \begin{subfigure}{.55\linewidth}
    \centering
    \includegraphics[width=\linewidth]{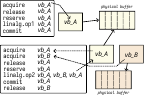}
    \caption{Virtual-buffer reuse.}
  \label{fig:vb-reuse}
  \end{subfigure}
  \hspace{3mm}
  \begin{subfigure}{.39\linewidth}
    \centering
    \includegraphics[width=\linewidth]{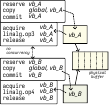}
    \caption{Physical-buffer reuse.}
  \label{fig:pb-reuse}
  \end{subfigure}
  \caption{Virtual-buffer optimizations.}
  \Description{Two diagrams illustrating reuse of a virtual buffer across exclusive handoffs and reuse of physical storage by virtual buffers with non-overlapping lifetimes.}
\end{figure}

\subsection{Reuse-Aware Virtual Buffer Concretization} \label{sec:buffer-reuse}
Because each \sysop{vb} requires a semaphore object and a memory region, both of which are physical resources, \sysname{} reduces their use to conserve hardware resources.
In particular, \sysname{} first reduces the number of virtual buffers when multiple tensor values can be represented by the same virtual buffer, and then assigns the remaining virtual buffers to physical storage and synchronization semaphores.

\paragraph{Reusing virtual buffers across exclusive handoffs.}
Keeping each virtual buffer distinct can be unnecessarily conservative.
\sysname{} therefore reuses virtual buffers when their dependences already form an exclusive handoff.
Many \lstinline{linalg} operations expose this structure naturally: an operation may update one of its operands in place, or the result of one operation may be consumed by a single downstream operation before that storage is reused.
In these cases, the connected virtual buffers can be merged into one buffer whose readable and writable state advances across the handoff.

Figure~\ref{fig:vb-reuse} illustrates this optimization.
The in-place update in \lstinline{linalg.op1} consumes and then rewrites the value carried by \texttt{vb\_A}, so the same virtual buffer can be released, reserved, and committed for the updated value.
The following \lstinline{linalg.op2} continues this handoff by consuming the committed value and writing its result back to the same virtual buffer.

\paragraph{Concretizing virtual buffers with reusable physical storage.}
After virtual-buffer reuse optimization, the remaining virtual buffers still describe the logical dependences required for pipeline-parallel execution, but \sysname{} has not yet materialized physical storage or synchronization resources.
Stage~3 materializes these virtual buffers, while reusing physical storage whenever liveness permits.
This step cannot be treated as a late target-lowering detail. Because \sysname{} derives local-storage usage from the virtual buffers and the backend's symbolic latency model treats local-memory capacity as a feasibility constraint on block sizes, unnecessary allocations can rule out larger blocks that would otherwise expose more operation-level parallelism.

During concretization, \sysname{} analyzes virtual buffer lifetimes and exclusive-handoff relations to determine which virtual buffers can safely share physical storage after previous accesses have completed.
\sysname{} groups shape-compatible virtual buffers, builds an interference graph for each group, and applies a coloring algorithm to assign non-interfering virtual buffers to the same \sysop{mem.alloc} result.
Thus, multiple virtual buffers may occupy the same physical storage at different program points, while each virtual buffer keeps its own synchronization resource for enforcing the access ordering expressed by its \sysop{vb.*} operations.

After concretization, the \lstinline{memref} views exposed by a virtual buffer are derived from its assigned \sysop{mem.alloc} storage, while the readable and writable state is implemented by the associated \sysop{sem.alloc} semaphore.
On the consumer path, \sysop{vb.acquire} and \sysop{vb.release} wait on and update this semaphore around the read view.
On the producer path, \sysop{vb.reserve} and \sysop{vb.commit} do the same around the write view.
Thus, concretization preserves the virtual-buffer interface used by earlier compiler passes while making the underlying storage accesses and synchronization effects explicit in the IR.

Figure~\ref{fig:pb-reuse} shows the distinct case of physical-storage reuse: \texttt{vb\_A} and \texttt{vb\_B} remain separate virtual buffers with independent synchronization state, but share one \sysop{mem.alloc} because their lifetimes do not overlap.








\subsection{Topology-Aware Intermediate Transformation}
For spatial dataflow architectures with a software-visible NoC and a homogeneous core array, how the SPMD grid is mapped onto the cores directly affects available parallelism, cross-core data reuse, and communication cost. \sysname{} therefore applies a topology-aware intermediate transformation to expand the dataflow-agnostic program into a pruned set of mapping candidates. It first enumerates topology-aware spatial mappings, then analyzes the cross-core reuse exposed by each retained mapping and rewrites eligible memory transfers into NoC-aware operations.

\paragraph{Topology-aware mapping.}

\modified{\sysname{} maps the logical SPMD grid onto the parallelism available on the target core topology, as specified by the core organization in the architecture description in Figure~\ref{fig:arch}. Rather than requiring a one-to-one correspondence between program axes and physical array dimensions, \sysname{} enumerates different logical views of the available cores, varying both how many cores participate and how their parallelism is organized across program axes. This flexibility is useful when parallelism is uneven across program axes. For example, when mapping a workload with a small batch dimension but a long sequence dimension onto a regular \(8\times8\) mesh, \sysname{} may logically view the array as \(2\times(4\times8)\), assigning 2-way spatial parallelism to batch and the remaining 32-way extent to sequence.

For each candidate, \sysname{} tiles the SPMD program grid over one legal view of the target core topology, following the same tiling practice as TileLoom~\cite{tileloom}. The spatial tile extents determine the cross-core parallel portion of each grid axis, while any remaining extent executes temporally.}

\paragraph{Spatial-reuse-aware pruning.}
\modified{The main compilation pressure in this framework is candidate-space growth: \sysname{} enumerates different logical views of the parallelism available on the target core topology and different placements of the program's parallel axes, with each retained mapping modeled and solved independently. Exhaustively considering every legal core view and every ordering of program axes can therefore become combinatorial. \sysname{} reduces this space before symbolic modeling by prioritizing mappings that expose both parallelism and cross-core data reuse.}

For each program axis, \sysname{} computes an axis score (\emph{ParallelExtent}, \emph{ReuseVolume}, \emph{ReuseAccessCount}). 
\emph{ParallelExtent} measures how much work can be distributed along the axis, while \emph{ReuseVolume} and \emph{ReuseAccessCount} characterize how much data can be shared when accesses are invariant along that axis and how frequently such reuse occurs. 


\modified{\sysname{} uses these scores to construct a dominance-based partial order over program axes: an axis is constrained behind another only when it is dominated in all three metrics; otherwise, both relative orders remain legal. Rather than enumerating all axis permutations, \sysname{} enumerates only the linear extensions of this partial order and combines them with the legal core views to form mapping candidates. This pruning retains the alternatives that are not dominated under \sysname{}'s reuse-and-parallelism heuristic while substantially reducing the candidate set passed to modeling and solving.}
By applying this pruning, \sysname{} eliminates up to 93\% of the legal candidate space for GEMM mappings, and Section~\ref{sec:eval} empirically evaluates the quality of the retained candidates.

\paragraph{Spatial reuse across cores.}
\modified{For each retained mapping candidate, \sysname{} derives the spatial communication pattern induced by its block-to-core placement.
It identifies cross-core reuse from affine global-memory access expressions: when an access is invariant along a spatial dimension, multiple cores address the same global-memory slice and may share the transferred data.
Subject to the connectivity and communication patterns supported by the target NoC, eligible accesses are rewritten as \sysop{copy} operations encoding the selected transfer mode and spatial sharing region.}
\begin{lstlisting}[style=mlir-druvbox-light-high-inline,numbers=none]
loom.copy %0, %1, @mem_global to @mem_local,
  multicast : [8, 1] region : (TL : [0, %2], BR : [7, %3])
\end{lstlisting}
\modified{Here, \texttt{multicast : [8, 1]} specifies the sharing extent, while \texttt{region} identifies the participating rectangular core region by its top-left and bottom-right coordinates.
Together with the underlying spatiotemporal mapping, each retained data-movement choice defines a schedule candidate. These annotations also provide the candidate-specific communication information used by the latency model to account for NoC transfer costs.}

\subsection{Latency Modeling}

\paragraph{Building the Hierarchical Execution Graph.}
A latency model for a SPMD program must expose where load, computation, and store stages may overlap.
\sysname{} captures this structure with a \emph{Hierarchical Execution Graph} (HEG), a recursively defined execution model:

\begin{align}
g &::= \mathsf{pipe}(\ell, g', s) \mid \mathsf{for}(i, g') \mid c \mid g; g \\
\ell &::= \mathrm{LOAD\_OP}^{*} \mid \emptyset \\
s &::= \mathrm{STORE\_OP}^{*} \mid \emptyset \\
c &::= \mathrm{COMPUTE\_OP}^{*}
\end{align}

\added{A sequential composition $g_1; g_2$ represents two sibling HEG subgraphs executed in order within the same scope, where $g_1$ and $g_2$ are arbitrary HEGs.}
A $\mathsf{pipe}$ node represents a load stage $\ell$, a child HEG $g'$, and a store stage $s$, where either memory stage may be empty.
The pipeline structure specifies a potential overlap; whether the overlap is enabled is left as a solver decision.
A $\mathsf{for}$ node repeats its child HEG $g'$ for $i$ iterations, where $i$ may be either a concrete integer or a symbolic expression.
A leaf HEG $c$ is a sequence of non-memory compute operations.
Figure~\ref{fig:heg} shows an example HEG. The root HEG consists of three subgraphs that are executed sequentially in this candidate. Its compute stage is a $\mathsf{for}$ node with child HEG $g'$ and trip count $i$, representing a loop nest whose body exposes pipeline-parallel execution.

\begin{figure}[htbp]
  \centering
  \includegraphics[width=\linewidth]{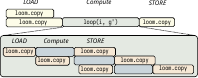}
  \caption{Example Hierarchical Execution Graph.}
  \Description{Hierarchical execution graph with sequential root stages and a nested loop whose load, compute, and store components may be pipelined.}
  \label{fig:heg}
  
\end{figure}

For example, Flash Decode naturally requires sequential composition: a local softmax loop can be represented as one HEG subgraph, followed by sibling subgraphs for gathering partial results, computing the global softmax, and applying the final score computation. Each of these subgraphs may itself contain nested loops or pipelines.

\paragraph{Primitive-based latency assembly.}
We use the term \emph{primitive operation} to refer to the lowest-level compiler-visible operation used to construct an HEG, such as a copy, a matrix multiplication, or an elementwise operation.
The architecture description module provides symbolic latency expressions for collections of such primitives using the architecture information in Figure~\ref{fig:arch}.
The HEG latency model treats these expressions as base terms: a leaf HEG is assembled from compute primitives, while each load or store stage is assembled from the corresponding memory primitives.
Thus, primitive performance modeling remains local to the architecture description module, while the HEG specifies how the resulting latency terms are composed.

Let $\mathcal{G}$ be the set of HEG nodes in a candidate program, and let $r \in \mathcal{G}$ denote the root HEG.
For each node $g \in \mathcal{G}$, $T_g(\mathbf{x}, \mathbf{b})$ denotes the symbolic latency expression of the subgraph rooted at $g$ under block-size configuration $\mathbf{x}$ and pipeline-knob configuration $\mathbf{b}$.

\added{For a leaf HEG $c$, the architecture description module directly provides its assembled compute latency $T_c(\mathbf{x})$.
For a sequential composition $g = g_1; g_2$, the two child subgraphs execute in order, so their latencies are added:
\begin{equation}
\label{eq:heg-seq}
T_g(\mathbf{x}, \mathbf{b})
=
T_{g_1}(\mathbf{x}, \mathbf{b})
+
T_{g_2}(\mathbf{x}, \mathbf{b}).
\end{equation}
}

For a $\mathsf{pipe}$ node $g=\mathsf{pipe}(\ell,g',s)$, the architecture description module provides the assembled memory-stage latencies $T_g^{\ell}(\mathbf{x})$ and $T_g^{s}(\mathbf{x})$ for the primitive collections in $\ell$ and $s$.
The remaining term, $T_{g'}(\mathbf{x}, \mathbf{b})$, is the latency of the child HEG and is assembled recursively from lower-level HEG nodes.
\sysname{} introduces a two-bit pipeline knob $\mathbf{b}_g=(b_g^{\ell m}, b_g^{ms})$ for this node.
The first bit controls whether the load--body boundary is opened as a pipeline boundary, while the second controls the body--store boundary.
A closed boundary fuses the adjacent components into one serial stage, whereas an open boundary allows them to execute as separate pipeline stages.
Therefore, the latency of a $\mathsf{pipe}$ node is:
\begin{equation}
\label{eq:heg-pipe}
\begin{aligned}
\ell_g &= T_g^{\ell}(\mathbf{x}), \quad
m_g = T_{g'}(\mathbf{x}, \mathbf{b}), \quad
s_g = T_g^{s}(\mathbf{x}), \\
T_g(\mathbf{x}, \mathbf{b}) &=
\begin{cases}
\ell_g + m_g + s_g, & \mathbf{b}_g = (0,0), \\
\max(\ell_g + m_g, s_g), & \mathbf{b}_g = (0,1), \\
\max(\ell_g, m_g + s_g), & \mathbf{b}_g = (1,0), \\
\max(\ell_g, m_g, s_g), & \mathbf{b}_g = (1,1).
\end{cases}
\end{aligned}
\end{equation}

For a $\mathsf{for}$ node $g=\mathsf{for}(i,g')$, \sysname{} composes the loop latency as:
\begin{equation}
\label{eq:heg-for}
T_g(\mathbf{x}, \mathbf{b})
=
i(\mathbf{x}) \cdot T_{g'}(\mathbf{x}, \mathbf{b}).
\end{equation}

Starting from leaf HEGs and memory stages, \sysname{} recursively assembles these expressions bottom-up.
The latency of the root HEG, $T_r(\mathbf{x}, \mathbf{b})$, is the total symbolic latency expression of the candidate program.

\added{\paragraph{Scope of symbolic latency modeling.}
This latency model targets regular SPMD grids, where block instances follow the same execution structure and can be described by a shared symbolic latency expression.
Irregular or asymmetric grids are more challenging because different grid positions may execute different amounts of work or follow different iteration extents, making their latencies position-dependent rather than representable by a single symmetric expression.
Such cases require either regularizing the iteration space before modeling or extending the latency model to explicitly represent position-dependent execution costs.}

\paragraph{Defining and solving the constrained optimization problem.}
The bottom-up HEG assembly gives each mapping candidate a symbolic latency expression.
For a retained candidate $\mathit{cdt} \in \mathcal{C}$, let $r_{\mathit{cdt}}$ denote its root HEG.
The expression $T_{r_{\mathit{cdt}}}(\mathbf{x}_{\mathit{cdt}},\mathbf{b}_{\mathit{cdt}})$ captures how the candidate latency is assembled.
Latency alone, however, is not sufficient to define a legal implementation.
The same architecture description module that provides primitive latency terms also provides resource and legality constraints, such as local-memory capacity, alignment requirements, divisibility constraints, and primitive validity conditions.
Together with candidate-specific placement and communication decisions, these constraints define a constrained optimization problem for each schedule candidate.

For each candidate $\mathit{cdt}$, \sysname{} solves:
\begin{equation}
\label{eq:heg-optimization}
\begin{aligned}
(\mathbf{x}_{\mathit{cdt}}^\star,\mathbf{b}_{\mathit{cdt}}^\star)
&=
\arg\min_{\mathbf{x}_{\mathit{cdt}},\mathbf{b}_{\mathit{cdt}}} T_{r_{\mathit{cdt}}}(\mathbf{x}_{\mathit{cdt}},\mathbf{b}_{\mathit{cdt}}) \\
&\quad \mathrm{s.t.}\quad \Phi_{\mathit{cdt}}(\mathbf{x}_{\mathit{cdt}},\mathbf{b}_{\mathit{cdt}}), \\[2pt] 
\Phi_{\mathit{cdt}}(\mathbf{x}_{\mathit{cdt}},\mathbf{b}_{\mathit{cdt}})
&:=
\left\{
\begin{aligned}
&x_k \in \mathbb{Z}_{>0}, \qquad x_k \le U_{\mathit{cdt},k}, \\
&x_k \equiv 0 \pmod{a_{\mathit{cdt},k}}, \\
&\mathrm{L1Footprint}_{\mathit{cdt}}(\mathbf{x}_{\mathit{cdt}},\mathbf{b}_{\mathit{cdt}}) \le C_{\mathrm{L1}}, \\
&\mathrm{ValidPrimitive}_{p,\mathit{cdt}}(\mathbf{x}_{\mathit{cdt}}), \\
&\mathrm{Legal}_{q,\mathit{cdt}}(\mathbf{x}_{\mathit{cdt}},\mathbf{b}_{\mathit{cdt}})
\end{aligned}
\right.
\end{aligned}
\end{equation}

After solving all retained candidates-specific optimization problems independently, \sysname{} selects the feasible candidate with the lowest modeled latency:
\begin{equation}
\label{eq:global-candidate-selection}
\mathit{cdt}^\star = \arg\min_{\mathit{cdt} \in \mathcal{C}_{\mathrm{feas}}}
T_{r_{\mathit{cdt}}}(\mathbf{x}_{\mathit{cdt}}^\star,\mathbf{b}_{\mathit{cdt}}^\star).
\end{equation}

The predicate $\Phi_{\mathit{cdt}}$ combines architecture constraints and program constraints under the fixed mapping candidate $\mathit{cdt}$.
The bounds $U_{\mathit{cdt},k}$ and alignment factors $a_{\mathit{cdt},k}$ come from the architecture description module combined with candidate-specific tiling context.
The L1 footprint constraint bounds the local-memory usage induced by the candidate, including additional buffering required by selected pipeline configurations.
Primitive validity constraints ensure that the latency terms requested from the architecture description module are applied within supported workload scenarios.
The remaining legality predicates capture semantic and candidate-specific requirements, such as valid communication patterns and tiling-bound constraints.
Infeasible candidates are discarded, and the best feasible solution is materialized as the selected mapping, block-size configuration, and pipeline configuration.

\added{
\subsection{Architecture Description}
\paragraph{Architecture-description requirements.}
\sysname{} separates kernel implementation from architecture-specific control. 
Kernel developers specify the operator without placement, communication, or pipelining decisions; \sysname{} infers these decisions jointly through the solver using the program structure and the target architecture description.
The architecture-specific effort is therefore concentrated in constructing this description, whose required information falls into two categories.
Structural properties, including core organization, interconnect topology and connectivity, and memory-hierarchy capacities, are obtained directly from hardware specifications. 
Performance properties, including the bandwidth of each memory level and the NoC, and the latency and throughput of primitive operations, are characterized using isolated microbenchmarks.
These parameters are collected once per architecture and reused across operators and input shapes, rather than measured within the per-kernel optimization loop.

\paragraph{Architecture-model applicability.}
\sysname{} targets dataflow architectures whose workload placement, communication, synchronization, and memory--compute pipelining are sufficiently explicit and predictable for static modeling. 
Its resource-based latency formulation is not tied to a homogeneous system or single-level memory hierarchy.
This abstraction naturally extends to heterogeneous functional units, hierarchical memories, and multi-level interconnects as long as their performance-relevant behavior can be modeled compositionally.
Less deterministic hardware behavior can instead be represented using conservative or offline-calibrated resource models, without introducing profiling into the per-kernel optimization loop.
}
\section{Evaluation} \label{sec:eval}
We evaluate \sysname{} on two generations of Tenstorrent accelerators. The evaluation answers four questions. 
First, can \sysname{} generate kernels that are competitive with Tenstorrent's optimized TTNN library and a tuning-based approach? 
Second, can the same architecture-parameterized compilation flow retarget multiple hardware generations?
Third, what compile-time cost does symbolic optimization introduce relative to tuning-based methods such as TileLoom?
Fourth, how accurately does the analytical model rank schedule candidates and choose block sizes?

\subsection{Experimental Setup}
\textbf{Platforms.}
We run all experiments on two servers: one with a Tenstorrent Wormhole card and one with a Tenstorrent Blackhole card. Table~\ref{tab:tt-specs} lists the host and accelerator specifications used in the evaluation.

\begin{table}[ht]
    \centering
    \caption{Specifications of the hardware platforms used in our evaluation.}
    \label{tab:tt-specs}
    \resizebox{.95\linewidth}{!}{
    \begin{tabular}{c|c|c}
    \hline
        & TT-Wormhole   & TT-Blackhole\\
        \hline
        Host CPU & Intel Xeon Silver 4514Y & AMD EPYC 7352 \\
        Host Memory & 256 GB & 512 GB \\
        OS & Ubuntu 24.04.4 LTS & Ubuntu 22.04.5 LTS \\
        \hline
        Topology       & 8 $\times$ 8  & 12 $\times$ 10 \\
        On-chip SRAM   & 108 MB        & 180 MB \\
        DRAM           & 12 GB GDDR6   & 32 GB GDDR6 \\
        Off-chip BW    & 288 GB/s      & 512 GB/s \\
        TFLOPS (FP16)  & 64            & 162 \\
        \hline
    \end{tabular}
    }
\end{table}

\textbf{Baseline and benchmarks.}
\added{We compare \sysname{} against two baselines: TTNN~\cite{tt-metal}, Tenstorrent's vendor-provided library of hand-optimized kernels for linear algebra and neural-network workloads~\cite{tt-metal-tech-reports}, and TileLoom~\cite{tileloom}, a state-of-the-art compiler for automatically generating optimized kernels on Tenstorrent hardware.}
We evaluate three representative kernels: GEMM, Flash Attention, and Flash Decode. These kernels cover different performance regimes. GEMM is a regular dense-linear-algebra primitive with structured tiling and high data reuse. Flash Attention combines matrix multiplication, softmax, and value aggregation, stressing fused dataflow, on-chip buffering, and communication scheduling. Flash Decode represents autoregressive inference, in which the query length is fixed and performance is dominated by streaming the KV cache. Together, these workloads test whether \sysname{} can generate competitive code across compute-bound, fused-attention, and memory-dominated kernels. 

\subsection{Kernel Performance}
\begin{table}[htbp]
\centering
\caption{Geometric-mean speedup over TTNN on Wormhole and Blackhole.}
\label{tab:overall_speedup}
\renewcommand{\arraystretch}{1.1}
\begin{tabular}{llcc}
\toprule
\textbf{Kernel} & \textbf{Method} & \textbf{Wormhole} & \textbf{Blackhole} \\
\midrule
\multirow{2}{*}{GEMM}
& TileLoom   & 0.950$\times$ & 1.100$\times$ \\
& \sysname{} & 0.989$\times$ & 1.160$\times$ \\
\midrule
\multirow{2}{*}{Flash Attention}
& TileLoom   & 1.940$\times$ & 1.980$\times$ \\
& \sysname{} & 2.293$\times$ & 2.151$\times$ \\
\midrule
\multirow{3}{*}{Flash Decode}
& TileLoom$^\ast$   & 0.840$\times$ & 0.870$\times$ \\
& \sysname{}$^\ast$ & 0.979$\times$ & 0.989$\times$ \\
& \sysname{} (all)  & 1.011$\times$ & 1.169$\times$ \\
\bottomrule
\multicolumn{4}{l}{\footnotesize
$^\ast$ Geometric mean over the subset of shapes supported by TileLoom.}
\end{tabular}
\end{table}

\added{
Table~\ref{tab:overall_speedup} reports the geometric-mean end-to-end speedup of TileLoom and \sysname{} over TTNN on Wormhole and Blackhole. A value greater than $1.0\times$ indicates a speedup over TTNN. \sysname{} consistently outperforms TileLoom across all three kernels and both platforms. For GEMM, \sysname{} achieves performance close to TTNN on Wormhole and a $1.16\times$ speedup on Blackhole. For Flash Attention, \sysname{} achieves $2.29\times$ and $2.15\times$ speedups on Wormhole and Blackhole, respectively, outperforming TileLoom's $1.94\times$ and $1.98\times$.

For Flash Decode, TileLoom does not support all evaluated shapes because of tool limitations. We therefore first compare both systems on the subset of shapes supported by TileLoom (marked with $^\ast$). On this common subset, \sysname{} improves the geometric-mean performance from $0.84\times$ to $0.98\times$ on Wormhole and from $0.87\times$ to $0.99\times$ on Blackhole. In contrast to TileLoom, \sysname{} supports all evaluated Flash Decode shapes; across the complete evaluation set, it achieves geometric-mean speedups of $1.01\times$ and $1.17\times$ over TTNN on Wormhole and Blackhole, respectively. All \sysname{} results are obtained out of the box, without per-shape profiling or profile-based platform-specific schedule tuning.
}

\modified{
    \subsubsection{GEMM}

We first evaluate GEMM, a compute-intensive primitive for which both TTNN and TileLoom provide strong baselines. Figures~\ref{fig:gemm_perf} and~\ref{fig:gemm_tileloom_perf} compare \sysname{} against the two systems across matrix dimensions from 256 to 16,384. On Wormhole, \sysname{} achieves $0.989\times$ the geometric-mean performance of TTNN and $1.050\times$ that of TileLoom. On Blackhole, \sysname{} improves over both baselines, reaching $1.160\times$ and $1.069\times$ geometric-mean performance relative to TTNN and TileLoom, respectively. The gains are particularly pronounced for several communication-sensitive shapes: for example, on Blackhole, \sysname{} reaches $1.50\times$ and $1.84\times$ the performance of TileLoom for $(16\mathrm{K},256,2\mathrm{K})$ and $(16\mathrm{K},256,16\mathrm{K})$, respectively.

\begin{figure*}[t]
    \centering
    \begin{subfigure}[t]{0.49\textwidth}
        \centering
        \includegraphics[width=\linewidth]{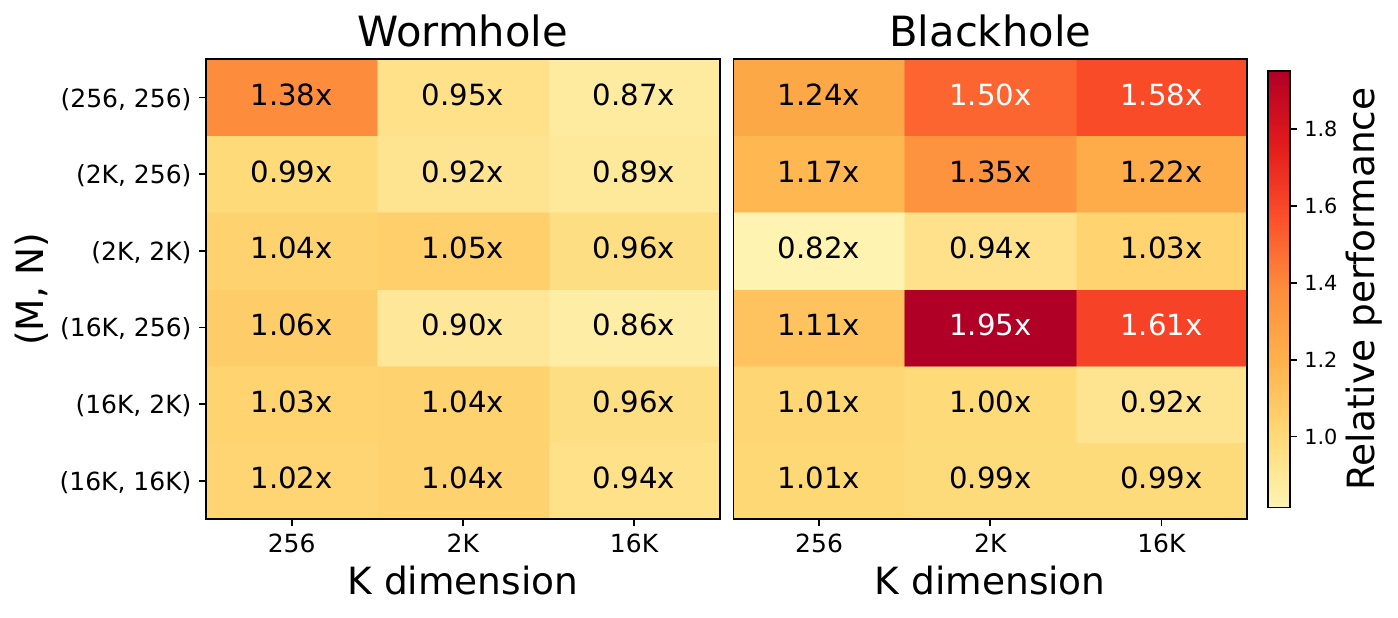}
        \caption{Performance relative to TTNN.}
        \Description{Heatmaps comparing GEMM performance of Loom and TTNN across
        matrix shapes on Wormhole and Blackhole.}
        \label{fig:gemm_perf}
    \end{subfigure}
    \hfill
    \begin{subfigure}[t]{0.49\textwidth}
        \centering
        \includegraphics[width=\linewidth]{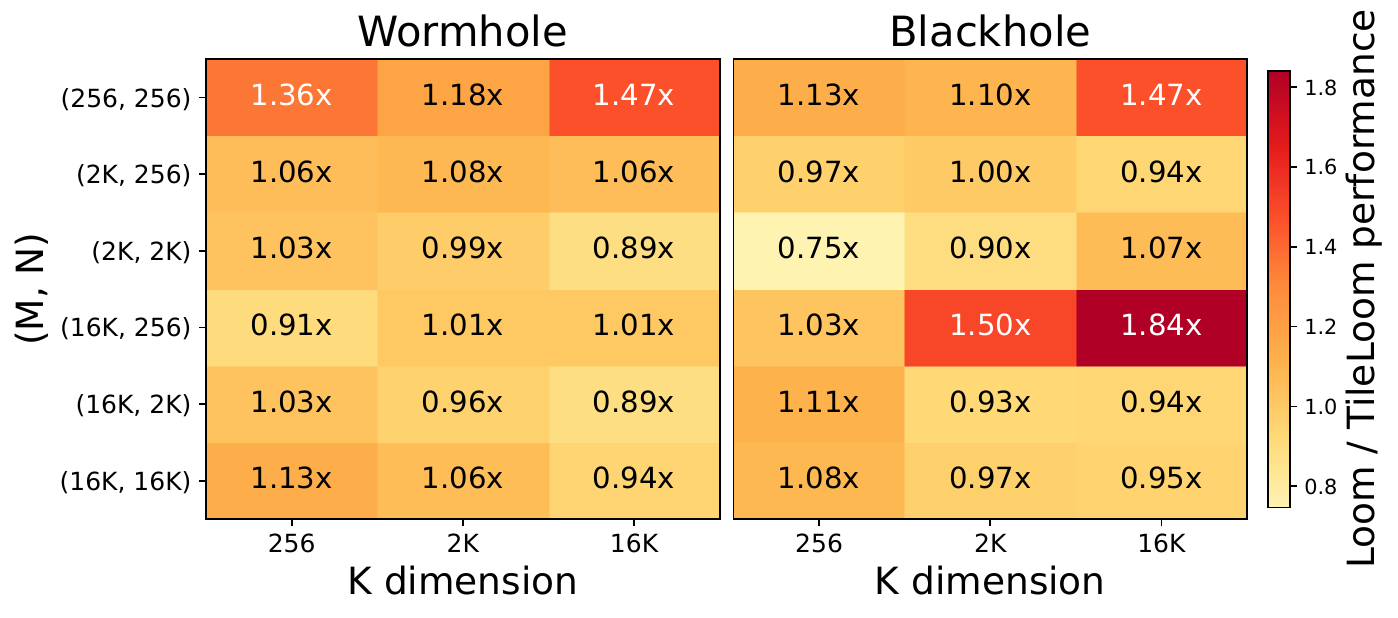}
        \caption{Performance relative to TileLoom.}
        \Description{Heatmaps comparing GEMM performance of Loom and TileLoom
        across matrix shapes on Wormhole and Blackhole.}
        \label{fig:gemm_tileloom_perf}
    \end{subfigure}
    \caption{GEMM performance of \sysname{} relative to TTNN and TileLoom on
    Wormhole and Blackhole.}
\end{figure*}

These results arise from three aspects of \sysname{}'s optimization flow. First, \sysname{} searches topology-aware data-movement schedules rather than restricting execution to a fixed set of manually designed templates. For example, on Blackhole, TTNN broadcasts one operand across the entire chip for $(16\mathrm{K},256,2\mathrm{K})$ and $(16\mathrm{K},256,16\mathrm{K})$. \sysname{} instead selects a two-dimensional broadcast that distributes the two operands along orthogonal directions, substantially reducing communication cost.

Second, \sysname{} jointly selects block sizes using its solver and latency model instead of relying on fixed tiling heuristics. This can improve performance even when the spatial communication pattern remains unchanged. For $(256,256,256)$, for example, \sysname{} selects a larger $K$ tile, amortizing data movement over more computation.

Third, \sysname{} co-optimizes tiling with intra-core pipelining. Double buffering can overlap communication with computation, but consumes additional L1 capacity and can therefore constrain tile size. \sysname{} explicitly captures this trade-off rather than enabling pipelining unconditionally. On Wormhole, for $(16\mathrm{K},16\mathrm{K},256)$, it disables double buffering and uses larger $M$ and $N$ tiles, increasing computation per data movement. Together, these choices allow \sysname{} to remain competitive with highly optimized TTNN kernels while also improving over TileLoom across both architectures.

\begin{figure*}[t]
    \centering
    \includegraphics[width=0.8\textwidth]{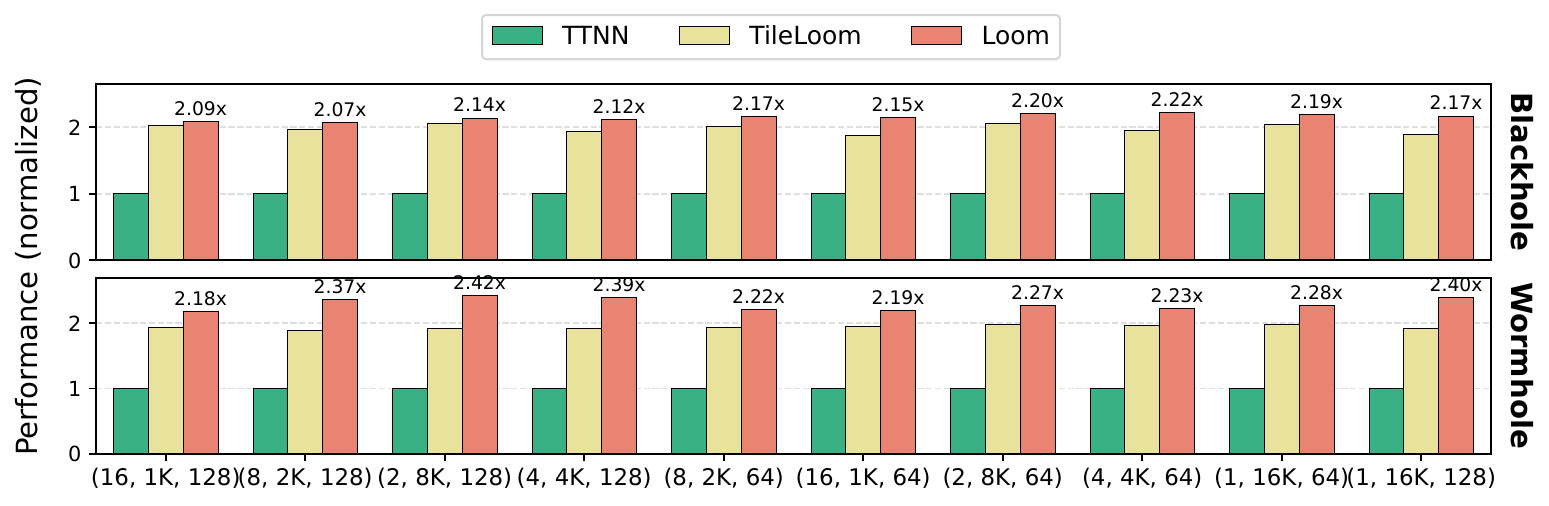}
    \caption{Performance of TileLoom and \sysname{} relative to TTNN on Flash Attention, with batch size $B$, sequence length $L$, and number of heads $N$.}
    \Description{Bar chart comparing the relative Flash Attention performance of TileLoom and Loom across input configurations on Wormhole and Blackhole.}
    \label{fig:flash_attention_perf}
\end{figure*}

Finally, the results demonstrate that the same optimization flow retargets across accelerator generations. \sysname{} parameterizes architectural properties including peak compute throughput, memory bandwidth, NoC cost, and local-memory capacity. Retargeting therefore requires updating the hardware model rather than redesigning GEMM schedules for each generation. The solver then adapts communication topology, tile sizes, and pipeline choices to the new hardware. This enables \sysname{} to move from near-TTNN performance on Wormhole to outperforming both TTNN and TileLoom on Blackhole without Blackhole-specific schedule engineering.
}

\subsubsection{Flash Attention}
Flash Attention is a more complex fused operator than GEMM. It combines matrix multiplication, softmax, and value aggregation, so performance depends on both compute efficiency and the ability to reuse data across cores while controlling communication. We evaluate the non-causal variant because it exposes more spatial dataflow choices and highlights the benefit of topology-aware mapping and NoC-aware memory rewriting.

\added{Figure~\ref{fig:flash_attention_perf} shows that \sysname{} consistently outperforms both TTNN and TileLoom on Wormhole and Blackhole. Across the evaluated configurations, \sysname{} achieves $2.18\times$--$2.42\times$ speedup over TTNN on Wormhole and $2.07\times$--$2.22\times$ on Blackhole. It also consistently improves over TileLoom, demonstrating the benefit of \sysname{}'s solver-based exploration in identifying better-performing implementations within the optimization space.}

\subsubsection{Flash Decode}
Flash Decode exercises a different regime from Flash Attention. In autoregressive decoding, the query length is one, and the kernel repeatedly streams over the KV cache. This makes the kernel more memory dominated: it has less dense compute reuse, and performance depends heavily on KV-cache movement. Because Flash Decode has limited parallel dimensions to distribute across cores, a key optimization opportunity is to parallelize the reduction dimension, i.e., the sequence-length dimension. Exploiting this opportunity requires careful inter-core communication planning together with an intra-core schedule that coordinates synchronization and overlaps communication latency. Flash Decode therefore stresses both inter-core data movement and intra-core pipeline optimization.

\added{Figure~\ref{fig:flash_decode_perf} reports the normalized performance of TileLoom and \sysname{} over TTNN for Flash Decode. Overall, \sysname{} consistently outperforms TileLoom and remains competitive with TTNN across the evaluated shapes. On Blackhole, \sysname{} outperforms TTNN on most shapes and achieves a speedup of up to $1.42\times$. On Wormhole, performance is more shape-dependent, with a peak speedup of $1.24\times$ and several configurations performing slightly below TTNN. In contrast, TileLoom is generally slower than TTNN on the supported shapes and reports N/A for several configurations on which it fails because of tool limitations.}

\begin{figure*}[htbp]
    \centering
    \includegraphics[width=0.8\textwidth]{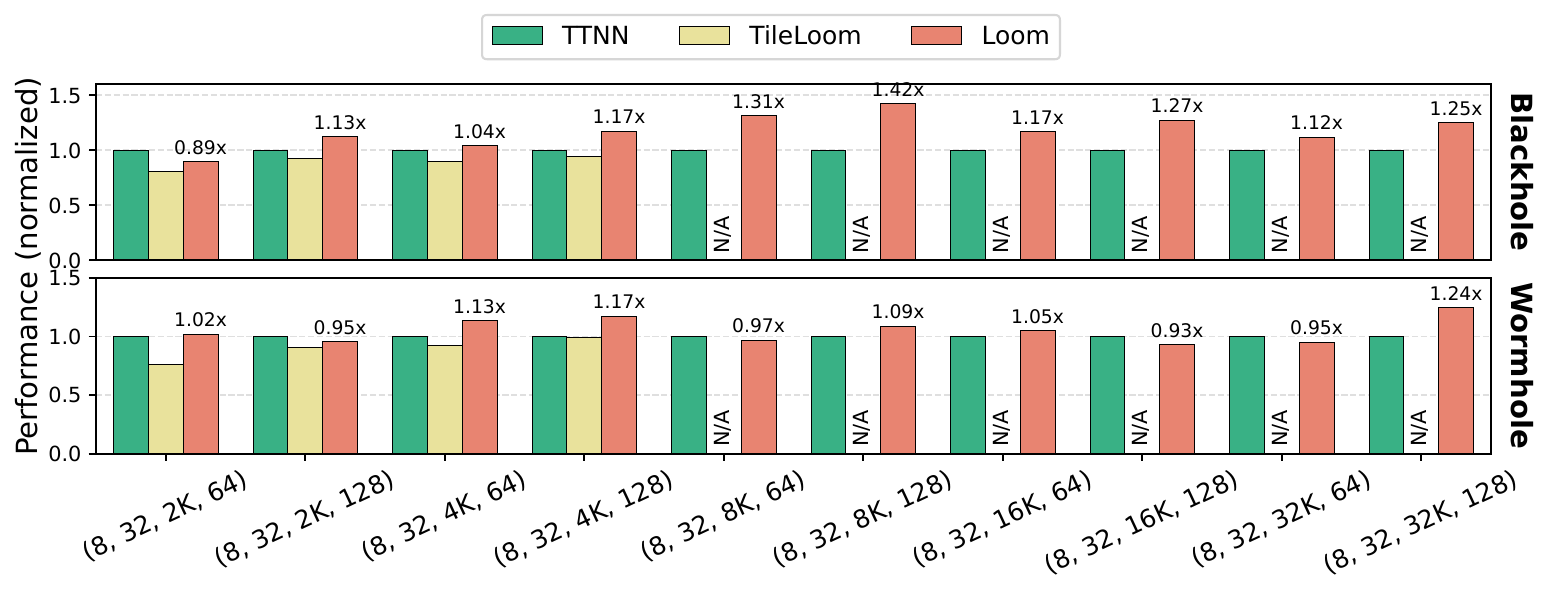}
    \caption{Performance of TileLoom and \sysname{} relative to TTNN on Flash Decode, with batch size $B$, number of heads $N$, sequence length $L$, and head dimension $D$.}
    \Description{Bar chart comparing the relative Flash Decode performance of TileLoom and Loom across input configurations on Wormhole and Blackhole.}
    \label{fig:flash_decode_perf}
\end{figure*}

This behavior is expected because Flash Decode offers less scheduling freedom. Since the query dimension is fixed to one, \sysname{} cannot use it as an effective spatial-mapping or tiling axis. The solver also has a smaller legal block-size space because the query tile is effectively predetermined. These constraints reduce the benefit of topology-aware mapping and symbolic tiling. Even under this restricted setting, \sysname{} remains close to or faster than TTNN, showing that the compiler can handle memory-dominated inference kernels without manual tuning.

\subsection{Compilation Cost}
\added{Table~\ref{tab:compilation_cost} compares the compilation cost of \sysname{} with that of TileLoom when generating the top-1 and top-5 candidates. The top-1 and top-5 searches take 6.74~s and 34.21~s, respectively, compared with 5.75~s and 27.66~s for TileLoom. Thus, although \sysname{} is 17.2\%--23.7\% more expensive, its compilation cost remains on the same scale as TileLoom's. In return, \sysname{} produces higher-quality candidates, improving the average speedup from $0.948\times$ to $0.989\times$ for the top-1 candidate and from $1.029\times$ to $1.089\times$ for the top-5 candidates, corresponding to 4.3\% and 5.8\% higher performance, respectively. \sysname{} therefore incurs a modest compile-time increase over TileLoom while producing higher-performing candidates in this experiment.}
\begin{table}[htbp]
\centering
\caption{Compilation cost and resulting performance of TileLoom and
\sysname{}.}
\label{tab:compilation_cost}
\small
\begin{tabular}{lcccc}
\toprule
& \multicolumn{2}{c}{\textbf{TileLoom}}
& \multicolumn{2}{c}{\textbf{\sysname{}}} \\
\cmidrule(lr){2-3}
\cmidrule(lr){4-5}
& \textbf{Top-1} & \textbf{Top-5}
& \textbf{Top-1} & \textbf{Top-5} \\
\midrule
Compilation Time (s) $\downarrow$
    & 5.75 & 27.66 & 6.74 & 34.21 \\
Avg. Speedup $\uparrow$
    & 0.948$\times$ & 1.029$\times$
    & 0.989$\times$ & 1.089$\times$ \\
\bottomrule
\end{tabular}
\end{table}

\subsection{Ablation Studies}
\subsubsection{Candidate ranking}

We next evaluate whether the latency model ranks spatial mappings accurately. Exhaustively running every schedule candidate with every legal block-size configuration is impractical because the combined search space is too large. Instead, we use a focused top-$5$ comparison. For each GEMM shape, we record the five schedule candidates ranked highest by the solver; for each candidate, the solver also selects the best legal block sizes and pipeline option under that candidate's constraints. We then run these five generated kernels on hardware and compare their measured performance.

Figure~\ref{fig:topk_candidates_perf} shows the measured performance range of the five candidates for each shape, normalized to the measured performance of the solver-ranked top-1 candidate. The horizontal line at $1.0$ is therefore the solver-selected candidate. If the interval ends at $1.0$, the solver-ranked top-1 candidate is also the measured-best candidate among the top five. If the interval extends above $1.0$, then another candidate in the top five is faster on hardware.

\begin{figure}[htbp]
    \centering
    \includegraphics[width=\linewidth]{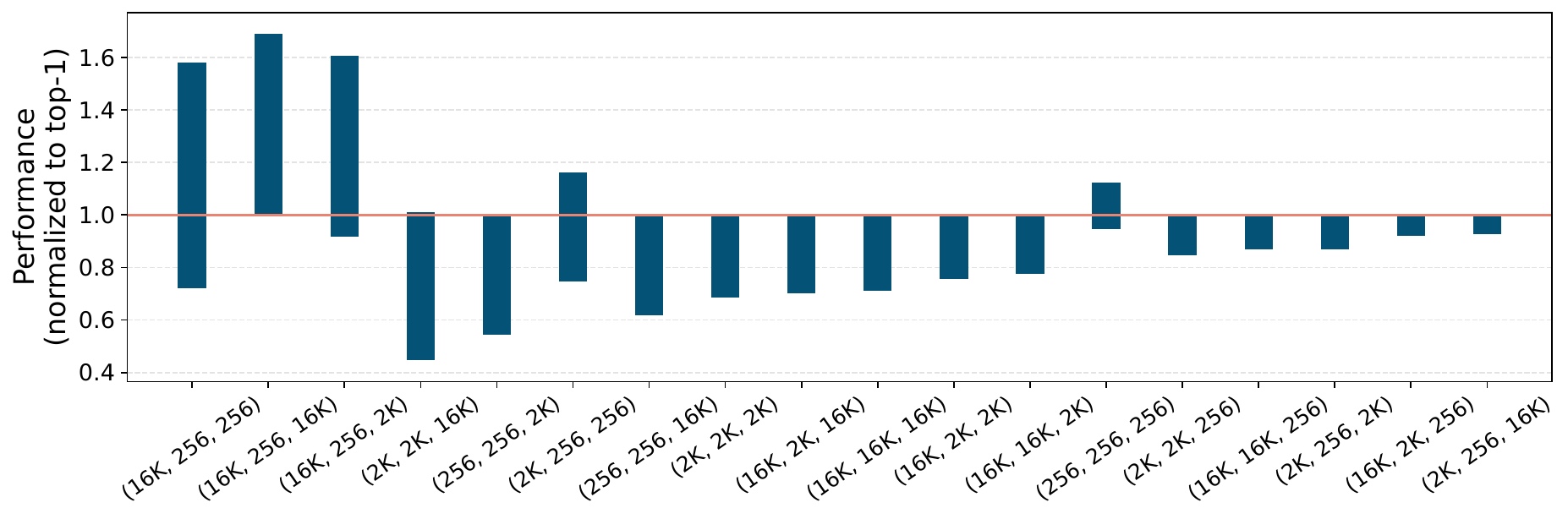}
    \caption{Measured performance range of the solver-ranked top-5 spatial-mapping candidates, normalized to the solver-ranked top-1 candidate.}
    \Description{Range plot showing measured performance among the five highest-ranked spatial-mapping candidates for each GEMM shape.}
    \label{fig:topk_candidates_perf}
\end{figure}

The results show that \sysname{} usually ranks the best mapping first. For most shapes, the upper end of the interval coincides with the $1.0$ line, meaning that the solver-selected candidate is also the best measured candidate among the top five. A few intervals extend slightly above $1.0$, but the gap is small, so the chosen candidate remains close to the best measured option. The largest deviations occur for the three leftmost, highly skewed shapes, for which the latency model does not rank the best measured top-5 candidate first.

These deviations are expected for highly skewed shapes, which provide too little parallel work along one or more matrix dimensions and make several spatial mappings nearly indistinguishable at the model level. Their measured ranking is then affected by low-level microarchitectural details that would require additional microbenchmarks and deeper hardware knowledge to model accurately. These effects explain the wider intervals, but the affected shapes are outliers. Moreover, even the optimized vendor library does not consistently handle these extreme shapes well, and our selected candidate remains close to, and sometimes faster than, the vendor library. The focus of this ablation study is to evaluate \sysname{}'s static modeling and scheduling capability, i.e., whether it can narrow a large search space to a small set of high-quality candidates without profiling every option. In practice, \sysname{} could add a final lightweight profiling step that runs the top-$k$ generated candidates on hardware and selects the best measured one as the final schedule.

\subsubsection{Block sizes}
We also evaluate whether \sysname{} selects effective block sizes after a spatial mapping has been fixed. For each GEMM shape, we keep the solver-selected candidate and treat its selected $M$, $N$, and $K$ block sizes as the center point. We then enumerate the local neighborhood by perturbing each block dimension to the nearest legal aligned value below, equal to, or above the solver choice. This produces up to $3^3=27$ configurations, including the solver-selected one. Configurations that violate hardware constraints are excluded.

As reported in Figure~\ref{fig:topk_block_sizes_perf}, each box summarizes the measured performance distribution of valid neighboring block-size configurations for a GEMM shape, normalized to TTNN.
The boxes show the 25th--75th percentile range with the median inside; the whiskers indicate the minimum and maximum, and the orange diamonds mark the configurations selected by the solver based on the latency expression.

\begin{figure}[htbp]
    \centering
    \includegraphics[width=\linewidth]{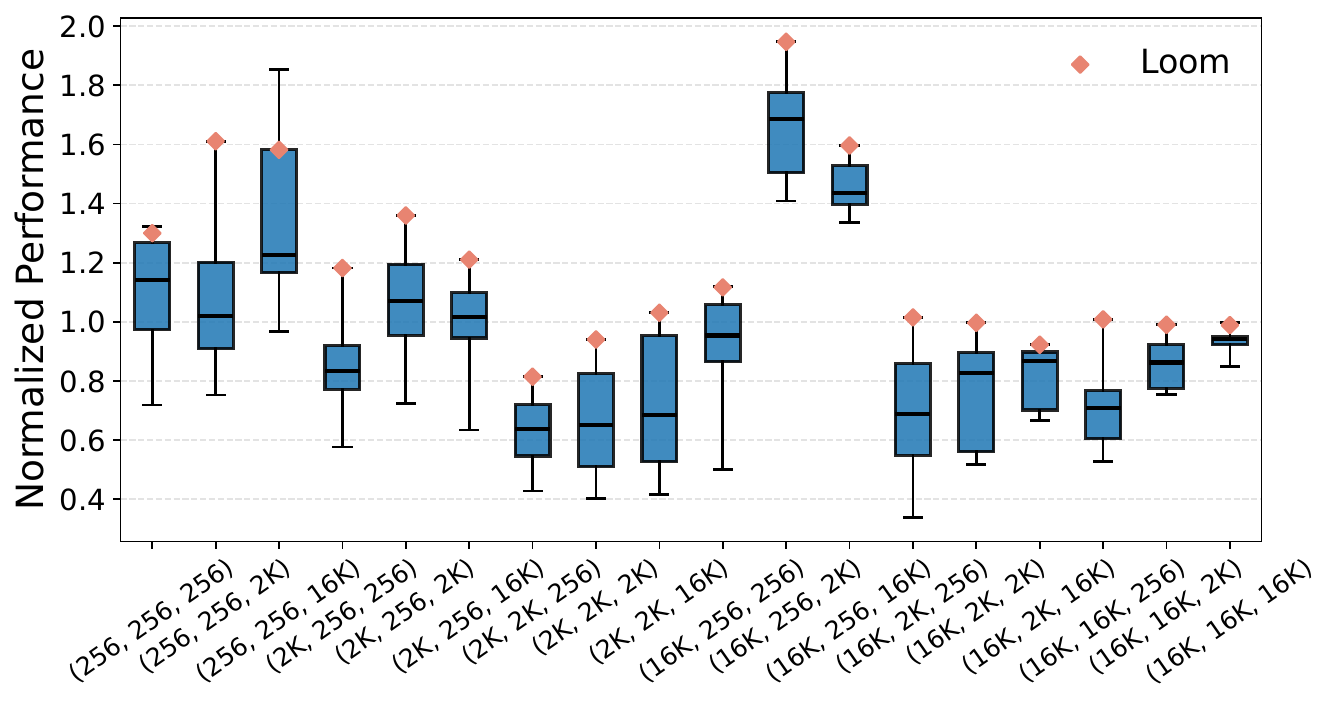}
    \caption{Performance over TTNN under local block-size perturbations around the solver-selected solution.}
    \Description{Box plots showing performance under nearby legal block-size choices, with the solver-selected configuration marked for each GEMM shape.}
    \label{fig:topk_block_sizes_perf}
\end{figure}

The solver-selected block sizes are among the best local choices for almost all GEMM shapes. In most cases, the orange diamond is at or near the top whisker, which means that local perturbations do not improve performance. Only a few shapes have a neighboring configuration that is slightly faster, and the gap is small. Thus, once the spatial mapping is fixed, the latency model usually identifies a block-size configuration that is locally optimal or very close to locally optimal on hardware.

This result matches the structure of GEMM. Under a fixed spatial mapping, the communication pattern is mostly determined, and the block-size choice mainly controls arithmetic intensity, local-buffer footprint, and primitive efficiency. Larger feasible blocks often improve compute utilization and push execution toward the compute-bound regime, and the symbolic latency model captures this trend. As a result, the solver often chooses the largest or near-largest feasible block sizes under capacity and alignment constraints. The rare mismatches occur for highly skewed shapes such as $(256,256,2\mathrm{K})$ and small shapes such as $(256,256,256)$, where there is too little work in some dimensions to reach stable throughput. In these regimes, memory bandwidth, compute utilization, and fixed overhead are harder for an analytical model to predict accurately.

\subsubsection{Model accuracy}
We validate \sysname{}'s model by comparing predicted and measured GEMM throughput. Figure~\ref{fig:cost_model} plots modeled and measured TFLOPS across $(M,N,K)$ configurations ranging from small matrices to large compute-heavy shapes. The goal is not cycle-accurate prediction. Instead, the model cheaply separates promising mappings from clearly inefficient ones during search.

\begin{figure}[htbp]
    \centering
    \includegraphics[width=\linewidth]{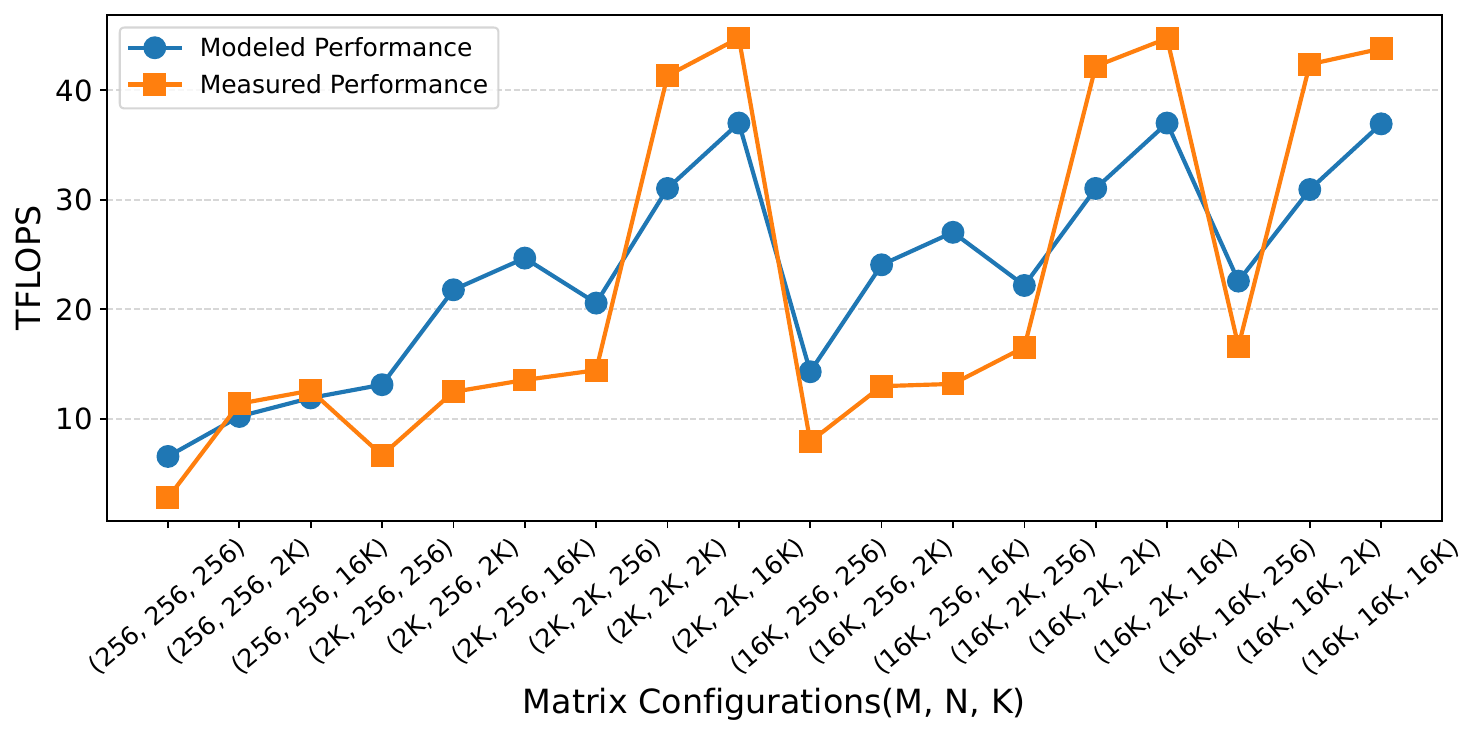}
    \caption{Accuracy of \sysname{}'s performance model against real hardware measurements.}
    \Description{Line chart comparing modeled and measured GEMM throughput across matrix shapes.}
    \label{fig:cost_model}
\end{figure}

The modeled curve preserves the main performance regimes observed on hardware. It predicts low throughput for small configurations, higher throughput for large compute-heavy configurations, and drops for shapes with limited parallelism or limited reuse. However, several gaps remain, including the transitions from $(2\mathrm{K},256,256)$ to $(2\mathrm{K},2\mathrm{K},256)$ and from $(16\mathrm{K},256,2\mathrm{K})$ to $(16\mathrm{K},2\mathrm{K},256)$. These differences arise from hardware effects that the model intentionally abstracts away, such as fixed kernel overhead, pipeline warm-up, synchronization cost, imperfect utilization, and the difficulty of reaching peak throughput on small or skinny matrices. Because the model uses idealized compute and memory throughput, it can overestimate configurations that do not fully utilize the accelerator.

Despite these absolute prediction errors, the model provides the information needed by the compiler. It narrows the search to high-quality regions rather than predicting the exact TFLOPS of every candidate. The candidate-ranking and block-size ablations in Figures~\ref{fig:topk_candidates_perf} and~\ref{fig:topk_block_sizes_perf} support this conclusion: the candidates selected by the model achieve competitive measured performance even when the modeled throughput differs from the hardware measurement. Thus, the model is accurate enough for search-space pruning and ranking.
\section{Related Work}

\paragraph{Analytical modeling for hardware--software co-design.}
Prior work on accelerator co-design and modeling, including Spatial~\cite{spatial}, Plasticine~\cite{plasticine}, AutoSA~\cite{autosa}, Timeloop~\cite{timeloop}, Accelergy~\cite{accelergy}, MAESTRO~\cite{maestro}, Sparseloop~\cite{sparseloop}, and AMOS~\cite{amos}, treats data movement, reuse, memory hierarchy, and hardware resource limits as explicit modeling objects.
These systems are primarily designed for hardware generation, accelerator design-space exploration, or mapping evaluation at the PE, buffer, dataflow-directive, or memory-hierarchy level.
\sysname{} adopts the same analytical view of data movement and resource constraints, but derives symbolic legality constraints and latency terms for compiling an executable tile program for a given sptial-dataflow target.

\paragraph{Compilation for spatial and tile-level programs.}
CGRA and FPGA compilers map dataflow graphs onto fixed fabrics through placement, routing, and scheduling~\cite{dsagen,mlcgra,mlircgra,morpher,casmap}, while spatial dataflow compilers such as TileLoom~\cite{tileloom}, Revet~\cite{revet}, LISA~\cite{lisa}, and system-level wafer-scale compilers~\cite{cerebras_compiler_mach} raise the abstraction to distributed cores, explicit communication, and architecture-aware placement.
Tile-level tensor compilers such as Triton~\cite{triton}, Ansor~\cite{ansor}, Hidet~\cite{hidet}, and TileLang~\cite{tilelang} expose blocked computation, local-memory movement, fusion, and pipeline structure, typically relying on tuning, learned cost models, or user-provided schedules.
\sysname{} is closest in spirit to these compilers, but keeps block sizes symbolic and jointly reasons about topology-aware mapping, NoC communication, local-buffer feasibility, and intra-core overlap instead of optimizing a fixed tile configuration or profiling concrete variants.

\paragraph{Symbolic and solver-based schedule optimization.}
\sysname{} is also related to analytical tile-size selection, polyhedral optimization, and solver-based mapping~\cite{solver_positivity,pluto,tiramisu,looper,satmapit}.
These approaches represent schedule choices as mathematical variables or search decisions under legality constraints.
\sysname{} applies this principle at the SPMD-block level for spatial dataflow programs, using symbolic variables for tiling factors and pipeline boundaries, and evaluating them with a hardware-parameterized analytical latency model.
The solver therefore serves as an interpretable bridge between the compiler IR and the hardware-derived performance model, rather than only as a search mechanism over a fixed schedule space.

\section{Conclusion}

This paper presents \sysname{}, a symbolic compiler framework for tile-based SPMD programs on spatial dataflow architectures.
By keeping part of schedule decisions symbolic, \sysname{} exposes the interaction among topology-aware placement, NoC communication, local-buffer capacity, and intra-core pipelining to an analytical performance model and solver.
This approach turns hardware constraints into explicit compiler reasoning rather than relying only on fixed templates, heuristics, or profile-based tuning.
Our evaluation on Tenstorrent Wormhole and Blackhole shows that \sysname{} can generate implementations that match and often outperform vendor-library kernels across GEMM, Flash Attention, and Flash Decode.
These results suggest that as accelerators continue to expose more software-controlled dataflow mechanisms, hardware-derived symbolic compilation can provide an interpretable and retargetable path toward high-performance code generation.

\bibliographystyle{ACM-Reference-Format}
\bibliography{references}
\end{document}